\documentclass[a4paper,12pt]{article}

\usepackage{graphicx}
\usepackage{ascmac}
\usepackage{leqno,here}
\usepackage{latexsym}
\usepackage{amsmath,amssymb}

	\newtheorem{theorem}{{\large T{\small{\sc HEOREM}}}}[section]
	\newtheorem{lemma}{{\large L{\small{\sc EMMA}}}}[section]
	
	\newtheorem{proposition}{{\large P{\small{\sc ROPOSITION}}}}[section]

\begin{document}

\begin{center}
{\Large On the maximum of Cram\'er's V}

\medskip
Etsuo Hamada

\medskip
{\scriptsize 
Faculty of Information Science, Osaka Institute of Technology, 

1-79-1 Kitayama, Hirakata, Osaka, Japan. 

}
\end{center}

\begin{abstract}
The Cram\'er's V is popular as an association coefficient 
in goodness-of-fit tests for contingency tables 
and its maximum value is known to be $1$, but it is not true. 
We propose a modified Cram\'er's V. 
\end{abstract}

\medskip
{\bf keyword:} Cram\'er's V

\section{Cram\'er's V}
\label{sec:CramerV}

The Cram\'er's V is popular as an association coefficient 
in goodness-of-fit tests for contingency tables. 
For a contingency table 

\medskip
\begin{table}[h]
\centering
\caption{Contingency table} \label{table:contingency}
\begin{tabular}{|c|*{5}{c}|c|} \hline
 & $B_1$ & $\cdots$ & $B_j$ & $\cdots$ & $B_c$ & sum \\
\hline
 $A_1$ & $x_{11}$ & $\cdots$ & $x_{1j}$ & $\cdots$ & $x_{1c}$ & $x_{1 \cdot}$ \\
 $\vdots$ & $\vdots$ &  & $\vdots$ &  & $\vdots$ & $\vdots$ \\
 $A_i$ & $x_{i1}$ & $\cdots$ & $x_{ij}$ & $\cdots$ & $x_{ic}$ & $x_{i \cdot}$ \\
 $\vdots$ & $\vdots$ &  & $\vdots$ &  & $\vdots$ & $\vdots$ \\
 $A_r$ & $x_{r1}$ & $\cdots$ & $x_{rj}$ & $\cdots$ & $x_{rc}$ & $x_{r \cdot}$ \\
\hline
 sum & $x_{\cdot 1}$ & $\cdots$ & $x_{\cdot j}$ & $\cdots$ & $x_{\cdot c}$ 
& $n$ \\
\hline
\end{tabular}
\end{table}

\medskip
the definition of the Cram\'er's V is 
\begin{equation} \label{eqn:cramer-V}
 {\rm V} \ = \ \sqrt{ {\chi^2 \over n \, \min(c-1,r-1)}}
\end{equation}
where $c$ is the number of columns, $r$ is the number of rows, 
and $\chi^2$ is the chi-square statistic of the contingency table as 
follows: 
\begin{equation} \label{eqn:chisq}
 \chi^2 \ = \ 
\sum_{i=1}^r \sum_{j=1}^c {(x_{ij} - e_{ij})^2 \over e_{ij}} ,
\end{equation}
where $e_{ij}$ is the expectation with respect to the observation $x_{ij}$. 
As the probability version for (\ref{eqn:chisq}), 
\cite{Cramer46} wrote, in page 282, that 
 
\medskip

\begin{center}
\begin{minipage}[h]{11cm}
{\small {\it 
On the other hand, by means of the inequalities 
$p_{ij} \leq p_{i \cdot}$ and $p_{ij} \leq p_{\cdot \, j}$ 
it follows from the last expression that $\varphi^2 \leq q-1$, 
where $q = \min(r,c)$ denotes the smaller of the numbers $r$ and $c$, 
or their common value if both are equal. }}
\end{minipage}
\end{center}

\medskip
\noindent
Note that symbols, etc. above are adapted to the format of this paper 
and the mean square contingency is 
\begin{equation} \label{eqn:varphi}
 \varphi^2 
= \sum_{i=1}^r \sum_{j=1}^c {(p_{ij} - p_{i\cdot} p_{\cdot j})^2 
  \over p_{i\cdot} p_{\cdot j}},
\end{equation}
and $\varphi^2 = \chi^2/n$ 
because of $x_{ij}= n \, p_{ij}$ and $e_{ij} = n \, p_{i\cdot} p_{\cdot j}$. 

\medskip
\begin{proposition}[\cite{Cramer46}] 
\label{prop:cramer}
$$
0 \ \leq \ {\varphi^2 \over q-1} 
 \ = \ {\varphi^2 \over \min(r,c) - 1} 
 \ \leq \ 1. 
$$
\end{proposition}

\medskip
The Cram\'er's V has since been defined as (\ref{eqn:cramer-V}) 
whose maximum value has been used as $1$. 
For example, see \cite{Akoglu18}, \cite{Okeka19}, and \cite{Urasaki24}.

\section{The maximum value of $V$}
\label{sec:maximum-value}

Since Cram\'er introduced the contingency coefficient $V$, 
its maximum value has been recognized as 
$n (\min(r,c) - 1)$, as in Proposition \ref{prop:cramer}.  
However, we recognize that this is a mistake and 
that the correct value is $n (rc - 1)$. 

\medskip
\begin{theorem} \label{thm:maximum}

For the $chi^2$ statistic (\ref{eqn:chisq}) of the contingency table, 
its maximum value is as follows: 
$$
 \max \chi^2 \ = \ n (rc - 1). 
$$
\end{theorem}

\begin{lemma} \label{lem:varphi}

For the mean square contingency (\ref{eqn:varphi}) of the contingency table, 
its maximum value is as follows: 
$$
 \max \varphi^2 \ = \ rc - 1. 
$$
\end{lemma}

\medskip
We first prove the Lemma. The proof of the theorem 
can be derived naturally from the result. 

\medskip
\noindent
{\bf (Proof of Lemma \ref{lem:varphi}) } \quad 
By using the relations $p_{ij} \leq p_{i \cdot}$ and 
$p_{ij} \leq p_{\cdot j}$ that \cite{Cramer46} showed, it holds that 
\begin{eqnarray*}
 \varphi^2 
&=& \sum_{i=1}^r \sum_{j=1}^c {(p_{ij} - p_{i\cdot} p_{\cdot j})^2 
  \over p_{i\cdot} p_{\cdot j}} \\
&=& \sum_{i=1}^r \sum_{j=1}^c {p_{ij}^2 \over p_{i\cdot} p_{\cdot j}} 
  - 2 \sum_{i=1}^r \sum_{j=1}^c p_{ij} 
  + \sum_{i=1}^r \sum_{j=1}^c p_{i\cdot} p_{\cdot j} \\
&=& \sum_{i=1}^r \sum_{j=1}^c {p_{ij}^2 \over p_{i\cdot} p_{\cdot j}} -1 \\
&\leq & \sum_{i=1}^r \sum_{j=1}^c {p_{i \cdot} p_{\cdot j} \over p_{i\cdot} p_{\cdot j}} - 1 
\ = \ \left( \sum_{i=1}^r \sum_{j=1}^c 1 \right)  - 1  \ = \ rc -1. 
\end{eqnarray*}
\hfill $\Box$

Thus we propose a modified Cram\'er's V as follows: 
\begin{equation} \label{eqn:modified-V}
 {\rm modified \ V} \ = \ \sqrt{ {\chi^2 \over n \, (c\, r-1)}}
\end{equation}

\section{Simulation}
\label{sec:simulation}

For two contingency tables, $2 \times 2$ and $3 \times 3$, for the total 
number $n=200$, we randomly generate the number of cells. 
We assume that the probabilities of the cells are all equal 
and that the number of simulations is 1000.

\begin{table}[t] 
\centering
\label{tab:compare}
\caption{simulation results for Cram\'er's V and a modified Cram\'er's V 
(the number of data is 200 and the number of simulations is 1000.) }
\begin{tabular}{|*{5}{c|}} \hline
 & \multicolumn{2}{c|}{$2 \times 2$ contingency table} & 
  \multicolumn{2}{c|}{$3 \times 3$ contingency table}  \\
\cline{2-5}
 & V  & modified V & V & modified V \\
\hline
Min. & 0.0837 & 0.0483 & 0.4774 & 0.2387 \\
1st Qu. & 0.7096 & 0.4097 & 1.038 & 0.519 \\ 
Median & 0.9358 & 0.5403 & 1.2656 & 0.6328 \\
Mean & 0.9815 & 0.5667 & 1.2874 & 0.6437 \\
3rd Qu.& 1.2369 & 0.7141 & 1.5263 & 0.7632 \\
Max. & 1.7321 & 1 & 2 & 1 \\
\hline
\end{tabular}
\end{table}

\setkeys{Gin}{width=0.45\textwidth}
\begin{figure}[t]
\includegraphics{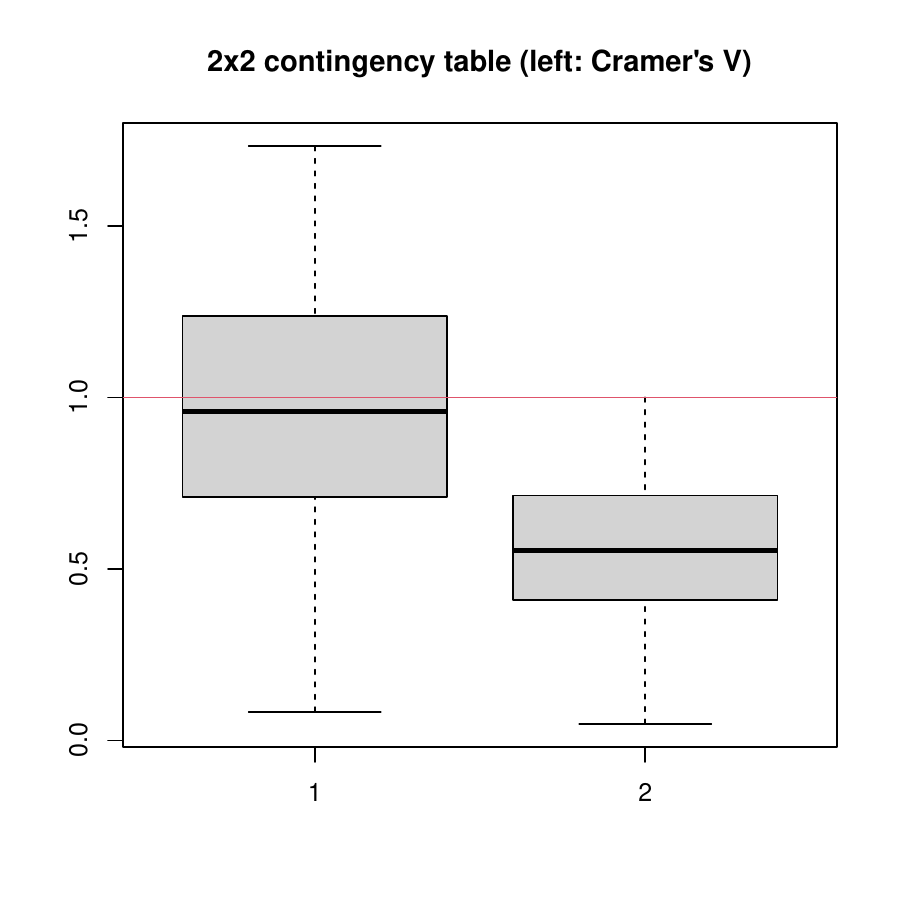}
\includegraphics{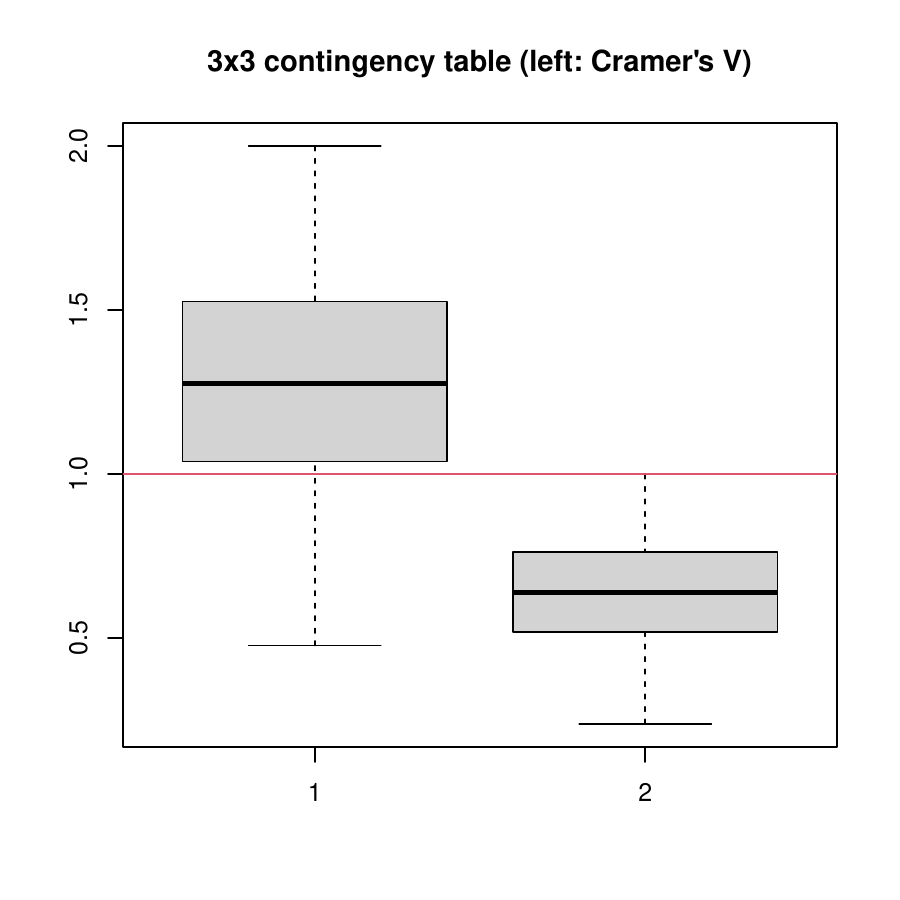}
\caption{simulation results for Cram\'er's V and a modified Cram\'er's V 
(the number of data is 200 and the number of simulations is 1000.) }
\label{fig:compare}
\end{figure}

\section{Conclusion}

The proof and simulation results regarding the maximum clearly 
show that we need to modify the Cram\'er's V. 
The previous contingency coefficients must 
be modified by using a modified Cram\'er's V with the maximum value $1$. 

\bigskip
{\bf Acknowledgment. } This paper was partially supported by 
Grant-in-Aid for Scientific Research (C) (general) 22K11946 
from the Ministry of Education, Culture, Sports, Science and 
Technology of Japan.

\end{document}